# Resonant Scanning with Large Field of View Reduces Photobleaching and Enhances Fluorescence Yield in STED Microscopy

Yong Wu[1,4,*], Xundong Wu[1], Rong Lu[1], Jin Zhang[1,3], Ligia Toro[1,3,4] and Enrico Stefani[1,2,4]

[1]Division of Molecular Medicine, Department of Anesthesiology, David Geffen School of Medicine, University of California, Los Angeles, CA, 90095, USA

[2]Department of Physiology, David Geffen School of Medicine, University of California, Los Angeles, CA, 90095, USA

[3]Department of Molecular & Medical Pharmacology, David Geffen School of Medicine, University of California, Los Angeles, CA, 90095, USA

[4]Cardiovascular Research Laboratory, David Geffen School of Medicine, University of California, Los Angeles, CA, 90095, USA

* ywu.thu@gmail.com

**Abstract**

Photobleaching is a major limitation of superresolution Stimulated Depletion Emission (STED) microscopy. Fast scanning has long been considered an effective means to reduce photobleaching in fluorescence microscopy, but a careful quantitative study of this issue is missing. In this paper, we show that the photobleaching rate in STED microscopy is slowed down and fluorescence yield is enhanced by scanning with high linear speed, enabled by the large field of view in our custom-built resonant-scanning STED microscope. The effect of scanning speed on photobleaching and fluorescence yield is more remarkable at higher levels of depletion laser irradiance, and virtually disappears in conventional confocal microscopy. With a depletion irradiance of >0.2 GW·cm$^{-2}$ (time average), we were able to extend the fluorescence survival time of the Atto 647N dye by ~80% with an 8-fold wider field of view. We confirm that STED Photobleaching is primarily caused by the depletion light acting upon the excited fluorophores. Experimental data agree with a theoretical model. Our results encourage further increasing linear scanning speed for photobleaching reduction in STED microscopy.

## Introduction

Stimulated Emission Depletion (STED) is a powerful technique in fluorescence microscopy that breaks the classical diffraction limit in a purely physical way. In STED microscopy, superresolution is achieved by adding a depletion laser beam with a doughnut-shaped focal spot, which is aligned with the excitation beam to inhibit the periphery fluorescence and record fluorescence only from the small central hole[1]. In theory, STED microscopy can reach unlimited resolution by indefinitely increasing the depletion laser irradiance. In practice, however, its resolution is limited by various factors, most notably severe photobleaching caused by the powerful depletion laser beam[2]. Therefore, in order to improve image quality and to enhance resolution for STED microscopy, reducing photobleaching is essential.

STED microscopy needs high depletion laser irradiance to reach high resolution. High excitation laser irradiance is often preferred as well to speed up the imaging process. Photobleaching has a linear dependence on weak excitation light irradiance with one-photon excitation (1PE), but for two-photon excitation (2PE) the dependence could have an order of three or higher[3, 4]. High excitation irradiance may also introduce nonlinearity even in 1PE[5]. In STED microscopy, the high power depletion pulses need to be stretched longer than 100 ps to avoid high-order photobleaching[2]. The triplet states play a crucial role in photobleaching[6, 7]. It was discovered that decreasing the pulsed laser repetition rate to <1 MHz greatly diminishes photobleaching of the Atto 532 dye. At the lower repetition rate, the fluorophores have enough time between consecutive pulses to relax from the triplet states, and fluorescence signal was increased by 5—25-fold[8]. Based on this discovery, the T-Rex (triplet relaxation) technique was invented, which can reduce photobleaching in STED microscopy with a tradeoff of slower imaging speed[9]. Using bunched laser pulses in T-Rex with suitable bunch duration can somewhat speed up imaging without losing the benefit of triplet state relaxation[10].

Another strategy to reduce triplet-state buildup is to adopt resonant scanning[11, 12]. For example, resonant scanning was applied to CW STED microscopy to reduce excessive photobleaching[13]. Photobleaching reduction due to fast scanning is nontrivial. A faster scanning speed results in shorter exposure time per scan, but the total exposure time is kept the same by accumulating more scans to reach the same, if not higher, level of fluorescence yield[11]. As previously reported, we built a STED microscope based on an 8 KHz resonant scanner that could reach ~40nm resolution in a 50 × 50 μm field of view (FOV)[14]. This large FOV increased the scanning speed by four times compared to the previous implementation[13]. We further expanded the microscope by adding a second channel and an ultrafast photon counting acquisition system[15]. In this paper, we use our custom-built STED microscope as a platform to quantitatively investigate the relationship between the linear scanning speed and the photobleaching rate. Although fast scanning is a conventional method to reduce photobleaching, as far as we know there is no careful study of this issue. Clarification of this relation is required to determine whether further increasing scanning speed is worthwhile.

## Theory

*Photobleaching measurement at different scanning speeds maintaining the same illumination dose*

When comparing the photobleaching rate at different scanning speed, we must maintain the same illumination dose in a unit area. In laser scanning confocal microscopy, image acquisition involves: 1) horizontal (x axis) scanning of the sample using a laser focal spot with laser irradiance $I$ and a linear speed $v$; 2) repetition of the horizontal scan for $L$ sequential lines in the vertical direction (y axis) with a constant spacing $\Delta y$ to generate a frame; and 3) the generation of the final image by a summation of $F$ frames. To maintain the same illumination dose in a unit area, we must have

$$F \cdot L \cdot I / v = \text{constant} \qquad (1)$$

With two different scanning speeds $v_1$ and $v_2$, there are two ways to satisfy Eq. (1): 1) maintaining the same laser irradiance, which we will call *equal-irradiance condition*, where $I_1 = I_2$ and $F_1 L_1 / v_1 = F_2 L_2 / v_2$;

and 2) maintaining the same total number of lines scanned referred as *equal-lines condition*, where $F_1 L_1 = F_2 L_2$ and $I_1/v_1 = I_2/v_2$.

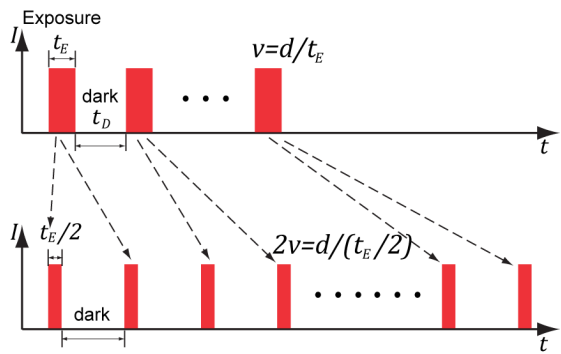

Fig. 1. Irradiance received by a fluorophore as a function of time under the equal-irradiance condition. At a linear scanning speed *v*, a fluorophore is exposed to illumination during exposure time-span $t_E = d/v$ (*d* is the focal spot diameter). When the speed doubles, the exposure time is reduced to $t_E/2$. To maintain the same illumination dose, the number of scans needs to be doubled.

We will mainly use the equal-irradiance condition to avoid the complication of nonlinearity[5]. Figure 1 shows light exposure under the equal-irradiance with a fast and a slow scanning speed. The plots illustrate the irradiance received by fluorophores as a function of time. At a scanning speed of *v*, in each line the fluorophore is exposed to light during a time-span (we call it *exposure time-span*) $t_E = d/v$, where *d* is the diameter of the laser focal spot (for simplicity we approximate the focal spot with a uniform circular disk). When the scanning speed is doubled, the same dose of illumination is split into two exposure time-spans, each of which has a duration of $t_E/2$. The two exposure time-spans are separated by another time-span without illumination, which we call the *dark time-span $t_D$*. In general, if the scanning speed is increased by a factor of *n*, the exposure time is shortened to $t_E/n$. We call *n exposure divisor*. To maintain the same illumination dose, when the scanning speed is increased by *n* times, the number of scans has to be increased by the same factor.

*Electronic states of a typical organic fluorophore: an analytic model*

The electronic states of a fluorophore can be modeled by a 5-state system, depicted in Fig. 2[5, 8]. We use $S^*$ to represent all the singlet states, and $T^*$ to include all the triplet states. Since the transitions within $S^*$ and $T^*$ states are much faster than the transitions between $S^*$ and $T^*$ (intersystem crossing and triplet-state relaxation) and the rates of photobleaching ($k_{bS}$ and $k_{bT}$), we can apply the steady-state approximation[5] where the relative populations of the inner states within $S^*$ and $T^*$ can be considered constant over time (Fig. 2, right panel). With this assumption we analytically solved the time dependence of the survival (unbleached) probability $R_{cont}(t)$ of fluorophores under quasi-continuous laser illumination (CW or high-repetition-rate pulsed lasers) (see Supplementary Information for deduction)

$$R_{cont}(t) = e^{-\beta t}\left(1 + \delta - \delta e^{-kt}\right) \qquad (2)$$

where *β* is related to the rate of photobleaching and *k* to the transition rates between $S^*$ and $T^*$, and *δ* is roughly the ratio between the triplet-state photobleaching rate and the intersystem transition rates. Since photobleaching is much slower than all other transitions, we have $\beta \ll k$, and $|\delta| \ll 1$. The parameters *β*, *k*, and *δ* are all derived from the rate constants in Fig. 2 (see Supplementary Information). Note that the excitation rate "constants" are actually dependent on the laser irradiance, possibly in a nonlinear manner due to the two-step photolysis[5].

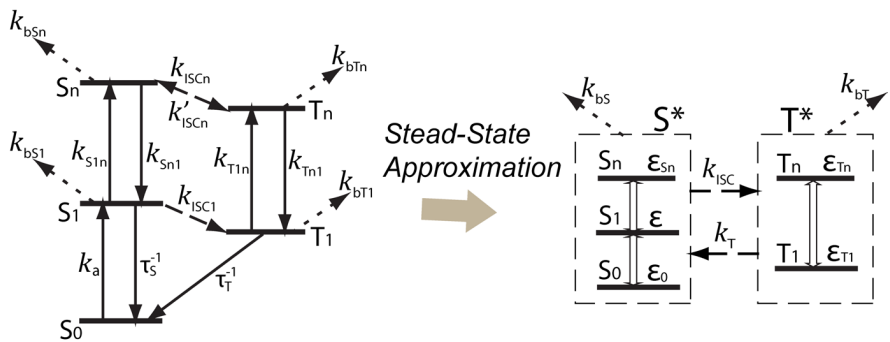

Fig. 2. Electronic states of a typical organic fluorophore. The ground state $S_0$ goes to the first excited state $S_1$ via light absorption. With a lifetime of $\tau_S$, relaxation from $S_1$ could produce fluorescence. $S_1$ could also undergo intersystem crossing (ISC) and enter the first triplet state $T_1$, which has much longer lifetime $\tau_T$ than $\tau_S$. Further excitations from $S_1$ and $T_1$ reach $S_n$ and $T_n$, respectively, which are connected by ISC and reverse ISC. Photobleaching could happen from all excited states. With the steady-state approximation, system reduced to 2 mixture-states $S^*$ (all singlet states) and $T^*$ (all triplet states). Within $S^*$ and $T^*$, the transitions are fast and the relative populations of states ($\varepsilon_0$, $\varepsilon$, and $\varepsilon_n$, in $S^*$; $\varepsilon_{T1}$ and $\varepsilon_{Tn}$ in $T^*$) are considered constant over time.

Eq. (2) is for quasi-continuous illumination. With fast resonant scanning, illumination is split into shorter exposure time-spans (see Fig. 1). When a single short exposure time-span $t_E$ is followed by a much longer dark time-span $t_D$ which is also much longer than the triplet-state lifetime, and when photobleaching resulted from the higher energy triplet state $T_n$ dominates over that from $T_1$[11, 16], the fluorophore survival probability after scanning for a time $t \gg t_E$ is

$$R_{scan}(t) = \exp\left(-\eta\left[\beta - \delta\left(1 - e^{-kt_E}\right)/t_E\right] \cdot t\right) \quad (3)$$

where $R_{scan}(t)$ decays exponentially with a time constant of $T_S = \eta\left[\beta - \delta\left(1 - e^{-kt_E}\right)/t_E\right]$, and $\eta = t_E/(t_E + t_D)$ is the scanning duty cycle. We call the time constant $T_S$ fluorescence survival time[17].

In a single frame, the fluorophore is scanned for $\sim d/\Delta y$ vertical lines. Recall that $d$ is the focal-spot diameter, and $\Delta y$ is the pixel height. In our experiments, $d \approx 300$ nm and $\Delta y = 15$ nm, and therefore a fluorophore is scanned for ~20 times in each frame. With an 8 KHz scanning mirror and bidirectional scanning, each cycle is 62.5 μs, and the dark time-span lasts for ~60 μs. If the fluorescent dyes have a much shorter triplet-state lifetime than 60 μs (e.g., the Atto 532 dyes has a triplet-state lifetime of ~1 μs[8]), then the population of the triplet states vanishes for each line.

The two common STED dyes under investigation in this paper can stay in the triplet states for milliseconds. In fact, Atto 647N was reported to have a triplet-state lifetime of 8–27 ms[18, 19], and Oregon Green 488 a triplet-state lifetime in milliseconds[20]. For these dyes, we consider a frame scan as one cycle. The dark time-span then depends on the acquisition frame rate, which is determined by the number of lines per frame. If each frame has 1000 lines, the dark time-span for each fluorophore is ~62.5 ms, which is long enough to relax the triplet states of our STED dyes. Considering the rounded shape of the focal spot, the exposure time-span is approximated by $t_E \approx \pi d^2/(4 \cdot v \cdot \Delta y)$. Although illumination within the exposure time-span is separated by ~60 μs gaps, it still can be considered as quasi-continuous due to the long triplet-state lifetime (in milliseconds), and Eq.(2) and Eq. (3) still hold.

Under the equal-irradiance condition, an *n*-fold increase in scanning speed divides $t_E$ into *n* equal portions, where *n* is the exposure divisor. After $t_E$, the fluorescence gain ratio due to the faster scanning speed is

$$\rho(n;\delta) = \frac{[1+\delta-\delta\exp(-kt_E/n)]^n}{1+\delta-\delta\exp(-kt_E)} \quad (4)$$

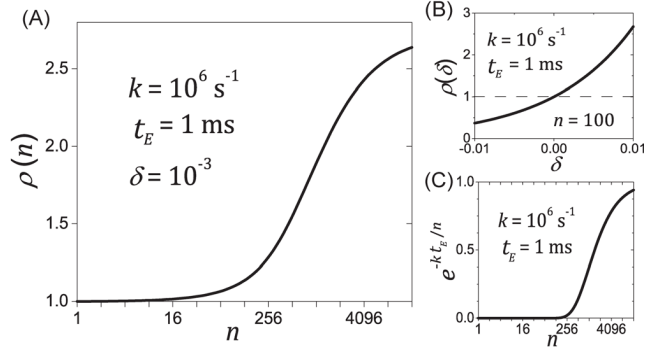

Fig. 3. Typical profile of $\rho(n;\delta)$, the fluorescence gain ratio through dividing exposure time-span $t_E$ into *n* divisions, as described by Eq. (4). (A) For given $\delta$, profile of $\rho(n)$ showing a slow entry, a quasi-linearly growth phase, and finally saturation. (B) For given *n*, $\rho(\delta)$ is a monotonically increasing function. (C) Profile of $\exp(-kt_E/n)$, which determines the profile of $\rho(n)$. Parameters used to plot the function are displayed in figures. Qualitative features of the $\rho(n)$ profile are not sensitive to parameter values.

Some features of $\rho(n;\delta)$ are: 1) fast scanning slows down photobleaching (i.e., $\rho(n;\delta)>1$) only when $\delta>0$, or equivalently, $k_{bT} > \varepsilon k_{bS}$ (see Fig. 2 and Supplementary Information). Intuitively, this is when the triplet-state photobleaching dominate over its singlet-state counterpart; 2) $\rho(n;\delta)$ is a monotonically increasing function of $\delta$ (Fig. 3B). Therefore, the effect of the scanning speed is more remarkable with growing $\delta>0$; and 3) for a given positive $\delta$, the profile of $\rho(n)$ is determined by $\exp(-kt_E/n)$. Fig. 3A illustrates the profile of $\rho(n)$, and Fig. 3C shows the profile of $\exp(-kt_E/n)$. The two profiles are very similar. The slow entry of $\rho(n)$ corresponds to the very small values of $\exp(-kt_E/n)$, the saturation happens when $\exp(-kt_E/n)$ approaches one, and in between is a quasi-linearly growing phase.

**Experimental Methods**

The experiments were conducted using our custom-built resonant-scanning dual-channel STED microscope with an ultrafast photon counting system[14, 15]. In brief, fluorescence excited by two pulsed/CW dual-mode lasers, with wavelengths of 635 nm and 488 nm (LDH-D-C-635 and LDH-D-C-485, PicoQuant, Germany), are depleted by a 750 nm pulse laser (Ti: Sapphire, Mai Tai HP, Newport, USA) and by a 592 nm CW fiber laser (2RU-VFL-P-2000-592, MPB Communications, Canada), respectively. Barrier filters were center at 669 nm (FF01-679/41, Semrock, USA) and 520 nm (FF03-525/50, Semrock, USA) for 635 nm and 488 nm lasers, respectively. To ensure excellent linearity of the acquisition system, photomultipliers (H7422-P, Hamamatsu Photonics) were used as detectors whenever possible. The frequency of the resonant scanner (CRS 8 KHz, Cambridge Technology, USA) was fixed and the scanning speed was controlled by changing the excursion of the horizontal resonant scanning mirror that determines the width of the field of view (FOV) and in turn the width of the image. As the frequency of the mirror is constant, a reduction of the excursion diminishes the FOV width and consequently the scanning velocity. The movement of the resonant scanner is sinusoidal with an almost linear portion at the center, and therefore we cropped the image around the center where the scanning speed was no less than 90% of the maximum speed at the image center. The maximum speed is given by $v_{max} = \pi \cdot W \cdot f$, where $W$ is the full FOV width, and $f = 8\,\text{KHz}$ is the frequency of the resonant scanner.

Table 1 lists the information of three different excursion of the horizontal mirror giving three different zooms (1, 4 and 8) and corresponding linear scanning speeds that we used in the experiments. We kept the same pixel size (15 × 15 nm) and the same number of lines in a frame for all zooms. As discussed in the previous section, to satisfy the equal-irradiance condition, the total number of frames taken at Zoom 1 was 8-fold as many as at Zoom 8, which resulted in 8-fold longer imaging time. On the other hand, under the equal-lines condition, the imaging time remains the same at different zooms and the excitation laser irradiance at Zoom 1 was 8-fold stronger than at Zoom 8. The same calculation was made for Zoom 4. The depletion laser irradiance was kept constant to maintain optical resolution.

We measured the photobleaching rate by taking a time series of images for the same FOV and recording the decay of image intensity (after subtracting a predetermined background caused by parasite light and dark counts). For each specimen, the fluorescence signal could vary greatly in different regions. Therefore, for each FOV, we normalized the image intensity to the image first taken. Each data point includes the mean plus the standard error calculated from at least 4 series of images in different regions. The illumination dose is measured by the normalized imaging time, which is defined as the actual imaging time to reach a certain illumination dose at Zoom 4 with predetermined laser irradiance. Therefore, by definition two experiments complete in equal normalized imaging times must receive the same dose of illumination. For example, under the equal-irradiance condition, a normalized imaging time of 10 seconds would mean 40 seconds of actual imaging at Zoom 1, or 5 seconds of actual imaging at Zoom 8.

Figure 4 shows the first and the second image, each with 3 minutes of imaging time (normalized), taken by the pulsed STED channel for Sample #2 (see Table 2). The figure illustrates the image sizes and the linear scanning speeds at different zooms and that the image intensity decay due to photobleaching is increased with higher zooms.

Varying the zoom from 1 to 8 by changing the FOV width with resonant scanning provided up to 8-fold change in linear scanning speed. To study the change in a wider range, we also employed a piezoelectric stage system (Nano PDQ 3-axis, controlled by Nano-Drive 85, 20 bit, Mad City Labs, USA) to access much slower scanning speeds. With the piezo-stage the scanning speed is independent of the FOV size. One controls the pixel size and pixel dwell time to determine the scanning speed. We kept the pixel dwell time at 1 millisecond in all experiments. Different pixel sizes were chosen to satisfy the equal-irradiance condition and the equal-lines condition, respectively (see Table 1). When the piezo-stage was used or compared to, an avalanche photodiode single photon counting module (SPCM-AQRH-12, Perkin Elmer, USA) was employed to detect fluorescence.

We studied two dyes that are common in STED microscopy: Atto 647N (Atto Tech, Germany) and Oregon green 488 (Life Technologies, USA). The former was excited and depleted by pulsed lasers (635 nm excitation and 750 nm depletion), whereas the latter was excited and depleted by CW lasers (485 nm excitation and 592 nm depletion). We used three fixed biological samples for our experiments. The information of the samples and the dyes is listed in Table 2. All specimens were mounted with ProLong® Gold (Life Technologies, USA) (see Supplementary Information for protocols).

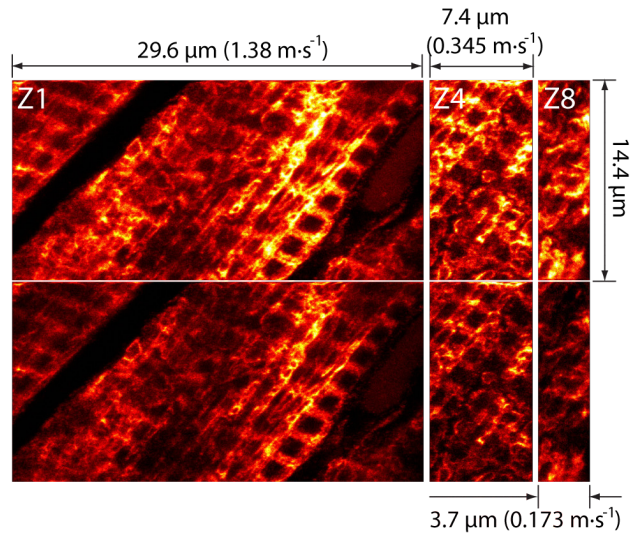

Fig. 4. Images taken at Zoom 1 (Z1), Zoom 4 (Z4) and Zoom 8 (Z8) for Sample #2. At all zooms, images have a pixel size of 15 × 15 nm. Images are cropped to only keep the portion with ≥90% of the maximum scanning speed. At Z1, the image width is 29.6 μm (1976 pixels). At Z4, the width is 7.4 μm (494 pixels), 1/4 of Z1. At Z8, the width is 7.4 μm (247 pixels), 1/8 of Z1. All zooms have the same height (14.4 μm; 960 pixels). Because the frequency of the resonant scanner is fixed, the scanning speed at Z1 (1.38 m·s$^{-1}$) is 4 times as fast as Z4 (0.345 m·s$^{-1}$), and 8 times as fast as Z8 (0.173 m·s$^{-1}$). The upper panels show images taken in 3 minutes of normalized imaging time, and the lower panels show images taken in the next 3 minutes. Image intensity decay due to photobleaching is 69% (Z1), 63% (Z4), and 53% (Z8).

We used the ImageJ (NIH, USA) and Origin 7.5 (Origin Lab, USA) software applications to visualize and analyze the data.

## Results

*Fast scanning significantly slows down photobleaching when depletion power is high*

We detected the fluorescence decay of Atto 647N in Sample #1 as a function of illumination dose with varying depletion laser irradiance. During the experiments, the 635 nm excitation laser beam had 0.22 mW optical power (throughout the paper, the laser power was measured at the back aperture of the objective), or ~300 kW·cm$^{-2}$ irradiance at the focal spot (in this paper, laser power or irradiance values are time-averages, unless stated otherwise). The depletion laser had up to 110 mW power, or up to ~0.4 GW·cm$^{-2}$ irradiance in focus. These values are what we usually use in practical STED imaging experiments.

Figure 5A shows the normalized image intensity decay as a function of the normalized imaging time in Sample #1. As expected, decay is faster with a higher depletion laser power. The lower zooms (faster scanning speeds) resulted in slower decay: the decay rate with 50 mW depletion power at Zoom 1 was almost the same as that with 10 mW depletion power at Zoom 8. This means widening the FOV by a factor of 8 would allow us to increase the depletion power by 5 times, which would reach much higher resolution without causing more photobleaching. Fig. 5B shows the ratio of the remaining fluorescence signal between Zoom 1 and Zoom 8 as a function of illumination dose. The ratio is always greater than one and increases with higher depletion laser power, demonstrating that the advantage of faster scanning becomes more significant at higher depletion laser irradiance. With 110 mW depletion power, the remaining fluorescence at Zoom 1 is up to ~235±46% as high as at Zoom 8 in 1 minute of normalized imaging time. In Fig. 5A, the data points were fitted to a mono-exponential decay function as expressed in Eq. (3), and the survival time (normalized) values are displayed in Fig. 5C. To quantify the difference in photobleaching rates at different zooms, the ratio of the survival time between at Zoom 1 and at Zoom 8 are plotted as a function of the depletion laser power in Fig. 5D. The ratio is always greater than one and increases with a growing depletion laser power. With 50 mW depletion power (irradiance is ~0.2 GW·cm$^{-2}$), the survival time at Zoom 1 is 95.5±1.1 seconds, while at Zoom 8 is 53.4±0.4 seconds. The ratio is thus ~1.79±0.02. In other words, an 8-fold higher scanning speed (resulted from an 8-fold wider FOV) can slow down the rate of photobleaching by ~80%.

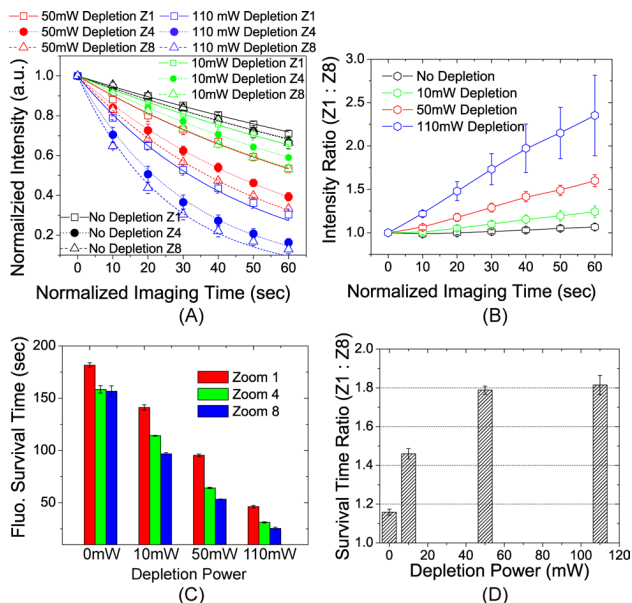

Fig. 5. Photobleaching rates in Sample #1 with varying depletion laser power. (A) Normalized Image intensity decay as a function of illumination (measured by normalized imaging time). Data points were fitted to mono-exponential decay expressed in Eq. (3). (B) Intensity ratio between Zoom 1 and Zoom 8 is always greater than one, and increases with higher illumination dose and growing depletion laser power. Lines are a guide for the eye. (C) Fluorophore survival time with different depletion laser power at different zooms. Lower zooms (faster scanning speed) have longer survival time. (D) Ratio of survival time between Zoom 1 and Zoom 8 is greater than one and increases with growing depletion power.

The same experiment was repeated for Sample #2 and the results are shown in Supplementary Fig. S1. Sample #2 is a rat heart tissue section, which is much brighter and bleaches much slower than Sample #1. The image intensity decay curves of this sample often do not fit well to mono-exponential decay, probably due to its higher fluorophore concentration[7] or more complex molecular environment. In spite of these differences, the relation between the scanning speed and the photobleaching rate in Sample #2 is qualitatively the same as in Sample #1, as illustrated in Supplementary Fig. S1. With 110 mW depletion power, Zoom 1 yielded up to ~80% more fluorescence than Zoom 8 in 1 minute of normalized imaging time (Supplementary Fig. S1B).

As shown in Fig. 5D, the effect of photobleaching slowing down by fast scanning is least notable when the depletion laser power is zero, i.e., for regular confocal microscopy. To confirm the effect was indeed related to depletion rather than a universal property only related to the photobleaching rate itself, we increased the excitation laser power in regular confocal condition and repeated the experiment. The results for Sample #1 are shown in Fig. 6. When the excitation laser power was increased to 0.44 mW (red curves in Fig. 6A), the photobleaching rates are comparable to 0.22 mW excitation + 50 mW depletion power at Zoom 8 (red dash curve in Fig. 5A). However, without depletion the effect of scanning speed on photobleaching rate is quite small. The decay curves were fitted to Eq. (3) and the ratio of the survival time between Zoom 1 and Zoom 8 are shown in Fig. 6B. The difference between two zooms is always less than 20%. Furthermore, with higher excitation power (hence faster bleaching rate), the ratio decreases. We also conducted the same experiment for Sample #2 (see Supplementary Fig. S2) and obtained the same result: the scanning speed has little impact on photobleaching rate when depletion is not present.

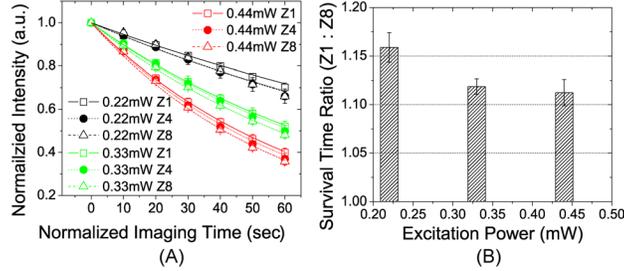

Fig. 6. Photobleaching in Sample #1 under regular confocal condition. (A) Image intensity decay curves fitted to Eq. (3). (B) Ratio of survival time between Zoom 1 and Zoom 8 as a function of excitation power. At high excitation power the photobleaching rate is comparable to that in STED microscopy. Fluorescence survival time is similar at all 3 zooms. Difference is at most ~15% with the lowest excitation power. At higher excitation power, the difference further diminishes.

In summary, the reduction of photobleaching by increasing the scanning speed is very pronounced when the depletion lasers power is high and it is minimal when using only the excitation laser (i.e., conventional confocal microscopy).

*STED photobleaching is primarily due to the depletion laser acting on the excited states*

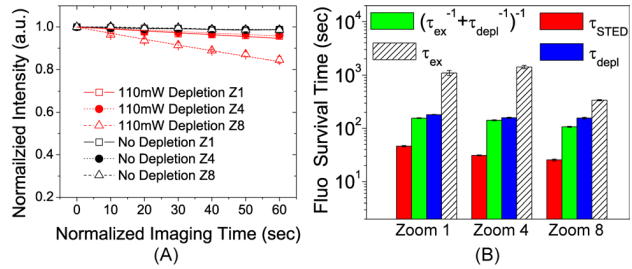

Fig. 7. (A) Image intensity decay caused by depletion only in Sample #1. Red curves are fitting to mono-exponential decay. Black curves represent background photobleaching caused by a low power (16 μW) excitation laser beam, which do not decay. (B) Comparison of STED survival time $\tau_{STED}$, excitation-only survival time $\tau_{ex}$, depletion-only time constant $\tau_{depl}$, and $\tau_{ind} = \left(\tau_{ex}^{-1} + \tau_{depl}^{-1}\right)^{-1}$ (would-be survival time if two lasers caused photobleaching independently). $\tau_{STED} \ll \tau_{ind} < \tau_{ex} < \tau_{depl}$ (note the logarithmic scale in B). See main text for details.

It was previously demonstrated that, when the duration of 760 nm depletion laser pulses is relatively long (~160 ps), they cause little photobleaching of the RH-414 dye, which primarily stems from multiphoton absorption[2]. Our pulsed depletion laser is at 750 nm with a pulse duration of ~400 ps, and we tested if the pulses would bleach the Atto 647N dye by themselves. The red curves in Fig. 7A show the fluorescence decay caused by a 110 mW depletion beam in Sample #1. The images were obtained with the excitation laser beam at very low power (16 µW), but the excitation beam was never turned on together with the depletion beam. The normalized imaging time shown in the horizontal axis is that of the depletion laser only. To measure the background photobleaching caused by the 16 µW excitation beam, the experiment was repeated with the depletion beam being blocked all the time and the results are shown by the black curves in Fig. 7A. The black curves do not show decaying, and therefore, the red decay curves were caused by the 110 mW depletion laser beam alone. In STED microscopy, if the depletion laser and the excitation laser caused photobleaching independently, one would expect its survival time to be $\tau_{ind} = \left(\tau_{ex}^{-1} + \tau_{depl}^{-1}\right)^{-1}$, where $\tau_{ex}$ and $\tau_{depl}$ is the survival time of the excitation-only bleaching and the depletion-only bleaching, respectively. From the black curves in Fig. 5A, the blue curves in Fig. 5A, and the red curves in Fig. 7A, the values of $\tau_{ex}$, $\tau_{depl}$ and the real STED survival time $\tau_{STED}$ were extracted and plotted in Fig. 7B (note the logarithmic scale in its vertical axis). It is obvious that $\tau_{STED} \ll \tau_{ind} < \tau_{ex} < \tau_{depl}$. Therefore, we concluded that the depletion laser must act on the fluorophores in the excited states to account for the photobleaching rates observed in STED microscopy. The same experimental procedure was repeated for Sample #2, and the results are shown in Supplementary Fig. S3. In Sample #2, the depletion caused a little fluorescence enhancement. Therefore, the above conclusion holds to be true.

*One cannot reduce STED photobleaching by simply using lower laser power*

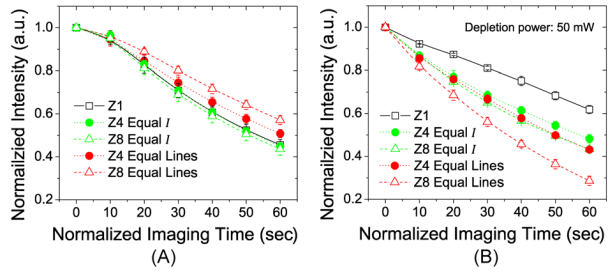

Fig. 8. Comparison of photobleaching under equal-lines and equal-irradiance (equal I) conditions for Sample #1. Zoom 1 data (black) is used as a reference. Green data points were recorded under equal-irradiance condition (i.e., using the same excitation irradiance as Zoom 1). Red data points were under equal-lines condition, using lower excitation irradiance. (A) In regular confocal condition, equal-lines condition has slower bleaching because lower laser power reduces nonlinear photobleaching. (B) In STED microscopy, the depletion power had to be kept the same as in reference to maintain resolution, and equal-lines condition has faster photobleaching because of extra depletion illumination. All lines are a guide for the eye.

As discussed in previous sections, there are essentially two ways to distribute a certain dose of illumination to a given FOV, equal-irradiance and equal-lines. Under the equal-lines condition, Zoom 8 should use 1/8 of the excitation irradiance as Zoom 1 uses. Due to two-step photolysis at higher excitation irradiance[5, 21], photobleaching has a super-linear (faster than linear) dependence on the excitation laser irradiance. Therefore, lower excitation irradiance should be beneficial to slow down photobleaching. In Fig. 8, we present the results of comparing photobleaching with different zooms under the equal-lines condition. In Fig. 8A, the black curve shows image intensity decay at Zoom 1 with an excitation power of 0.22 mW, which was used as reference. The red curves are decay at Zoom 4 and Zoom 8 under the equal-lines condition. Contrary to that under the equal-irradiance condition (green curves), the lower zooms with lower excitation irradiance are more favorable. In Fig. 8B, we present the results in STED microscopy. Because the optical resolution of STED microscopy is determined by the depletion irradiance, we maintained the same depletion laser power for all zooms, and hence at Zoom 8 fluorophores received an 8-fold as high as the depletion illumination dose than at Zoom 1. In other word, we had to break the equality of illumination dose to keep the equality of optical resolution. At higher zooms, the effect of the higher depletion illumination dose competes against the benefit from the lower excitation irradiance. The red curves in Fig. 8B show that, the higher depletion illumination dose dominates and faster scanning speed is overall beneficial.

*CW STED results*

We repeated the above experiments for the Oregon green 488 dye used in Sample #3, which was excited and depleted by CW lasers, and most of the results are qualitatively the same as for the Atto 647N dye. In Supplementary Fig. S4A and Supplementary Fig. S4B, one can see that the faster scanning speed resulted in slower photobleaching, and that such effect is emphasized by higher depletion power. With an 8:1 scanning speed ratio, we could get up to ~39±9% more fluorescence in 1 minute of normalized imaging time. Survival time was extracted (Supplementary Fig. S4C) by fitting the decay curves to Eq. (3) (plus a constant background). At Zoom 1, the survival time is 45±0.6 seconds, ~45±2% longer than at Zoom 8 (31±0.4 seconds) with 220 mW depletion power (Supplementary Fig. S4D). Supplementary Fig. S5 demonstrates that, this effect does not exist in regular confocal condition, even when the excitation power was so high that the photobleaching rate was comparable to STED. Similar to Fig. 8, Supplementary Fig. S6 illustrates photobleaching under the equal-lines condition (red curves): in regular confocal condition, the lower zooms had less photobleaching; in STED, it was the other way around because more depletion illumination dose was needed to maintain the optical resolution. The only major difference of Sample #3 is that the 592 nm depletion laser could cause significant photobleaching by its own, and more so at a higher zoom, as shown in Supplementary Fig. S7A. But STED photobleaching is still much quicker than that caused by the excitation laser and the depletion laser acting separately in time (Supplementary Fig. S7B). It again supports the theory that STED photobleaching is primarily due to the depletion laser affecting the excited fluorophores.

*Piezo-stage scanning results*

Scanning speed of the piezo-stage is much slower than the resonant scanner (see Table 2). Under the equal-irradiance condition, Eq. (1) dictates that the pixel size of the piezo-stage ought to be 110 × 110 nm. This size would be too large for a practical STED imaging experiment. The equal-lines condition is what a practical STED imaging experiment would use, under which the pixel size was chosen to be the same as in resonant scanning (15 × 15 nm), the excitation laser power was reduced to 3.9 µW, and the depletion power was maintained the same to maintain the optical resolution. In Fig. 9, we show the photobleaching comparison of piezo-stage and resonant scanning. Due to the slowness of piezo-stage, we only took two images for each time series to measure photobleaching (each image with 3 minutes of normalized imaging time). Fig. 9A shows the normalized image intensity of the second image to quantify photobleaching. Under the equal-irradiance condition, though the exposure time of the piezo-stage is 100-fold longer than resonant scanning at Zoom 8, their photobleaching rates are about the same. Piezo-stage scanning under the more practical equal-lines condition caused much more severe photobleaching because of excessive depletion illumination dose to maintain the optical resolution. For the four equal-irradiance cases, using the piezo-stage scanning case as the reference, the fluorescence gain $\rho(n)$ as a function of exposure divisor n is plotted in Fig. 9B. Its profile is consistent with the function profile depicted in Fig. 3. Piezo-stage scanning and Zoom 8 resonant scanning together show the slow entry, while Zoom 4, Zoom 2 and Zoom 1 belong to the quasi-linearly growth phase of the $\rho(n)$ profile. Similar results were obtained for Sample #2 and Sample #3, as illustrated in Fig. 9C and Fig. 9D, respectively. The experimental data points were fitted to the theoretical model expressed by Eq. (4), and the fitted parameter values are shown in Fig. 9 caption.

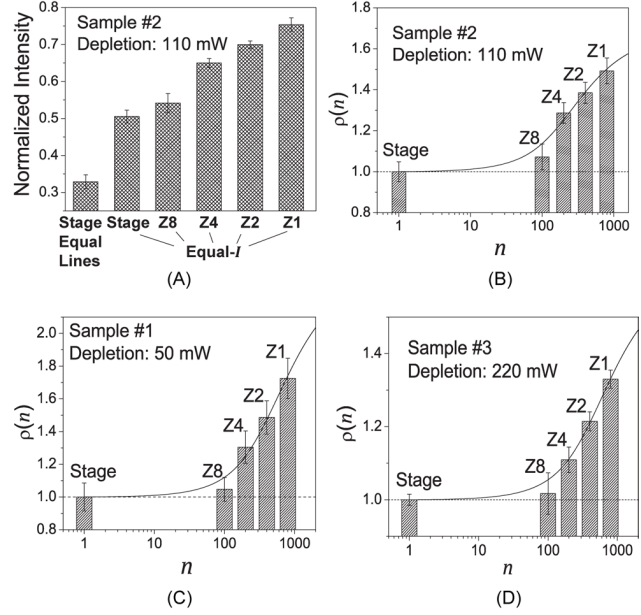

Fig. 9. Photobleaching rates of piezo-stage and resonant scanning in STED microscopy. (A) Normalized image intensity of the 2nd image in the time series for Sample #2. Under equal-irradiance condition (pixel size 110 × 100 nm), piezo-state caused about the same photobleaching as resonant scanning at Zoom 8, despite its exposure time-span is 100-fold longer. Under equal-lines condition (pixel size 15 × 15 nm), piezo-stage resulted in much more severe photobleaching. (B), (C), and (D) show fluorescence gain ratio $\rho(n)$ with respect to piezo-stage scanning as a function of exposure divisor $n$, under equal-irradiance condition for Sample #2, Sample #1 and Sample #3, respectively. Data points (columns) were fitted to $\rho(n)$ expressed in Eq. (4) (solid lines), which give $k = 2.7(\pm 0.99) \times 10^7$ s$^{-1}$; $\delta = 1.15(\pm 0.2) \times 10^{-3}$ (Sample #1); $k = 1.4(\pm 0.4) \times 10^7$ s$^{-1}$; $\delta = 1.33(\pm 0.24) \times 10^{-3}$ (Sample #2); and $k = 3.2(\pm 0.5) \times 10^7$ s$^{-1}$; $\delta = 5.4(\pm 0.4) \times 10^{-4}$ (Sample #3).

All the experiments discussed above were designed to resemble practical STED imaging and thus the depletion beam had a doughnut-shaped cross-section. But the zero-intensity center complicates the photobleaching process. We repeated one of the experiments (110 mW depletion power, Sample #2) with a Gaussian depletion beam and compared the result to standard STED in Supplementary Fig. S8. The image intensity curves under two conditions only have minor differences, suggesting that the complexity induced by the doughnut-shaped depletion beam is not very significant.

**Summary and Discussion**

The main conclusion of this paper is that the linear scanning speed has a nontrivial impact on the photobleaching rate and fluorescence yield in resonant scanning STED microscopy. Since the frequency of the resonant scanner is fixed, the linear scanning speed can be readily controlled by changing the width of the scan FOV. Using two types of organic dyes in three fixed biological samples, we have shown that an 8-fold wider FOV can extend the fluorescence survival time by up to 80% at high depletion irradiance. When the depletion power is low, this impact diminishes. For regular confocal microscopy, it may disappear altogether at the photobleaching rates comparable to that in STED microscopy. Photobleaching in STED microscopy is more severe than that caused by separate illumination of the depletion laser and the excitation laser, suggesting that the primary mechanism of STED photobleaching requires the fluorophores first being excited by the excitation laser.

Higher depletion irradiance emphasizes the effect of scanning speed on photobleaching, because it more efficiently depletes the excited singlet state population. As discussed in the *Theory* section, $\rho(n;\delta)$ increases with growing $\delta > 0$. With higher depletion irradiance, fluorescence suppression is higher and thus the population of the excited states $\varepsilon$ decreases ($\varepsilon \downarrow$). For our dyes with a triplet-state lifetime in milliseconds, $\delta \approx k^{-1}(k_{bT} - \varepsilon k_{bS})$ and most likely $\delta \uparrow$. Therefore, $\rho(n;\delta) \uparrow$ and the effect of scanning speed becomes more notable. On the other hand, when there is no depletion and the excitation power increases, $\varepsilon \uparrow$ is likely to cause $\rho(n;\delta) \downarrow$, and the effect diminishes, as illustrated in Fig. 6B.

Our experimental results indicate that a faster scanning speed is not useful in conventional resonant scanning confocal microscopy. Instead of increasing the scanning speed by *n* times, one should lower the excitation irradiance by $1/n$, because photobleaching has a super-linear dependence on the excitation power. However, this approach does not apply to STED microscopy, because we cannot lower the depletion irradiance; otherwise the optical resolution is compromised.

We have shown that resonant scanning is far more superior than piezo-stage scanning (with 1 ms pixel dwell time) in terms of photobleaching reduction, mainly because fluorophores in piezo-stage scanning receive excessive depletion illumination (it had to use the equal-lines condition to reach a pixel size of 15 × 15 nm). Piezo-stage scanning under the equal-irradiance has roughly the same photobleaching rate as at Zoom 8. It indicates that Zoom 8 ($t_E \approx 27\mu s$) just enters the quasi-linearly growth phase of Eq. (4). Therefore, further increase of linear scanning speed by using resonant scanner with a higher frequency (e.g., SC30 16 KHz, Electro-Optical Products Corp., USA) and wider FOV width is very likely to save us more fluorescence signal from photobleaching.

Increasing the scanning speed will shorten the pixel dwell time and raise the pixel clock rate. In our current system, the pixel clock rate is ~112.5 MHz at Zoom 1. We have built an ultrafast photon counting system that provides up to 450 MHz pixel clock rate, which allows for further scanning speed boost[15]. Another limit to viable scanning speed is the concern of *motion blur*. If the optical resolution at rest is 50 nm, Zoom 1 currently has a blurred resolution of ~53 nm in the horizontal direction. If the scanning speed is further doubled (quadrupled), the horizontal resolution would become ~59 nm (~73 nm)[15]. For the fluorescent dyes with milliseconds of triplet lifetime, one could use interlaced scan (laser switched on for one line and off for the next) to shorten the exposure time-span without changing the physical scanning speed and thus avoid motion blur.

Resonant scanning STED microscopy with high-repetition-rate pulsed lasers is similar to the bunched T-Rex technique[10], in the sense that a fluorophore would encounter the same temporal pattern of illumination: illumination bunches in bunched T-Rex are equivalent to exposure time-spans with resonant scanning. However, with resonant scanning one fluorophore's dark time is another fluorophore's exposure time, whereas in bunched T-Rex the dark time is universal for all fluorophores. Therefore, resonant scanning is more efficient in time. In particular, the T-Rex technique can hardly use fluorophores with a triplet-state lifetime in milliseconds, because the imaging speed would be too slow.

**Acknowledgement**

This work was supported by NIH R01 HL088640 (ES) and NIH R01 HL107418 (ES & LT).

**Author Contributions**

YW designed and conducted the experiments, developed the theory, and wrote the main manuscript text. XW designed and constructed the image acquisition system. RL and JZ prepared the biological sample and wrote the protocols. LT and ES provided research funds and revised the manuscript. ES conceived the idea of this study.

**Additional Information**
Competing financial interests: The authors declare no competing financial interests.

Reference List

1. Hell,S.W. *et al.* Nanoscale Resolution with Focused Light: STED and Other RESOLFT Microscopy Concepts in *Handbook of biological confocal microscopy* (ed. Pawley,J.B.) 571-579 (Springe, New York, 2006).

2. Dyba,M. & Hell,S.W. Photostability of a fluorescent marker under pulsed excited-state depletion through stimulated emission. *Applied Optics* **42**, 5123-5129 (2003).

3. Chen,T.S., Zeng,S.Q., Zhou,W., & Luo,Q.M. A quantitative theory model of a photobleaching mechanism. *Chinese Physics Letters* **20**, 1940-1943 (2003).

4. Patterson,G.H. & Piston,D.W. Photobleaching in two-photon excitation microscopy. *Biophysical Journal* **78**, 2159-2162 (2000).

5. Eggeling,C., Widengren,J., Rigler,R., & Seidel,C.A.M. Photobleaching of fluorescent dyes under conditions used for single-molecule detection: Evidence of two-step photolysis. *Analytical Chemistry* **70**, 2651-2659 (1998).

6. Song,L.L., Varma,C.A.G.O., Verhoeven,J.W., & Tanke,H.J. Influence of the triplet excited state on the photobleaching kinetics of fluorescein in microscopy. *Biophysical Journal* **70**, 2959-2968 (1996).

7. Song,L.L., Hennink,E.J., Young,I.T., & Tanke,H.J. Photobleaching Kinetics of Fluorescein in Quantitative Fluorescence Microscopy. *Biophysical Journal* **68**, 2588-2600 (1995).

8. Donnert,G., Eggeling,C., & Hell,S.W. Major signal increase in fluorescence microscopy through dark-state relaxation. *Nature Methods* **4**, 81-86 (2007).

9. Donnert,G. *et al.* Macromolecular-scale resolution in biological fluorescence microscopy. *Proceedings of the National Academy of Sciences of the United States of America* **103**, 11440-11445 (2006).

10. Donnert,G., Eggeling,C., & Hell,S.W. Triplet-relaxation microscopy with bunched pulsed excitation. *Photochemical & Photobiological Sciences* **8**, 481-485 (2009).

11. Borlinghaus,R.T. MRT letter: high speed scanning has the potential to increase fluorescence yield and to reduce photobleaching. *Microsc. Res. Tech.* **69**, 689-692 (2006).

12. Tsien,R.Y. & Bacskai,B.J. Video-rate confocal microscopy in *Handbook of biological confocal microscopy* (ed. Pawley,J.B.) 459-478 (Springer, New York, 1995).

13. Moneron,G. *et al.* Fast STED microscopy with continuous wave fiber lasers. *Optics Express* **18**, 1302-1309 (2010).

14. Gardeazabal Rodriguez,P.F. *et al.* Building a fast scanning stimulated emission depletion microscope: a step by step guide in *Current Microscopy Contributions to Advances in Science and Technology* (ed. Mendez-Vilas,A.) 791-800 (Formatex Research Center, Extremadura, 2012).

15. Wu,X., Toro,L., Stefani,E., & Wu,Y. Ultrafast Photon Counting Applied to Resonant Scanning STED Microscopy. *Journal of Microscopy* **Epub ahead of print**, (2014).

16. Deschenes,L.A. & Bout,D.A.V. Single molecule photobleaching: increasing photon yield and survival time through suppression of two-step photolysis. *Chemical Physics Letters* **365**, 387-395 (2002).

17. Wennmalm,S. & Rigler,R. On death numbers and survival times of single dye molecules. *Journal of Physical Chemistry B* **103**, 2516-2519 (1999).

18. Vogelsang,J. *et al.* A reducing and oxidizing system minimizes photobleaching and blinking of fluorescent dyes. *Angewandte Chemie-International Edition* **47**, 5465-5469 (2008).

19. Ha,T. & Tinnefeld,P. Photophysics of Fluorescent Probes for Single-Molecule Biophysics and Super-Resolution Imaging. *Annual Review of Physical Chemistry, Vol 63* **63**, 595-617 (2012).

20. Folling,J. *et al.* Fluorescence nanoscopy by ground-state depletion and single-molecule return. *Nature Methods* **5**, 943-945 (2008).

21. Eggeling,C., Volkmer,A., & Seidel,C.A.M. Molecular photobleaching kinetics of rhodamine 6G by one- and two-photon induced confocal fluorescence microscopy. *Chemphyschem* **6**, 791-804 (2005).

**Tables**

**Table 1. Scanning conditions used in experiments**

|  | FOV size (µm)[a] | Pixel size (nm) | Linear scanning speed (m·s$^{-1}$) | Exposure time per scan $T_E$ (µs) |
|---|---|---|---|---|
| **Zoom 1** | 29.6 × 14.4 | 15 × 15 | 1.38 | 3.4 |
| **Zoom 4** | 7.4 × 14.4 | 15 × 15 | 0.345 | 13.7 |
| **Zoom 8** | 3.7 × 14.4 | 15 × 15 | 0.173 | 27.3 |
| **Pizeo-stage equal-irradiance** | Any | 110 × 110 | 1.1×10$^{-4}$ | 2,730 |
| **Pizeo-stage equal-lines** | Any | 15 × 15 | 1.5×10$^{-5}$ | 20,000 |

[a] The portion of FOV with ≥90% of maximum scanning speed is shown.

**Table 2. Information of biological samples used in experiments**

|  | **Dye** | **Cell/tissue type** | **Primary antibody** | **Excitation laser** | **Depletion laser** |
|---|---|---|---|---|---|
| **Sample #1** | Atto 647N | Hela cell | Anti-ADP/ATP carrier | 635 nm, ~150 ps pulses, repetition rate 80 MHz | 750 nm, ~400 ps pulses, repetition rate 80 MHz |
| **Sample #2** | Atto 647N | Rat heart tissue section | Anti-Cytochrome c | 635 nm, ~150 ps pulses, repetition rate 80 MHz | 750 nm, ~400 ps pulses, repetition rate 80 MHz |
| **Sample #3** | Oregon green 488 | Mouse tissue section | Anti-VDAC | 485 nm, CW | 592 nm, CW |

# Resonant Scanning with Large Field of View Reduces Photobleaching and Enhances Fluorescence Yield in STED Microscopy

## Supplementary Information

*Deduction of Equations*

With the steady-state approximation, we only need to consider a 2-state system consisting of $S^*$ and $T^*$, whose population $P_S$ and $P_T$ satisfy

$$\begin{cases} \dfrac{dP_S}{dt} = -\left(\varepsilon k_{ISC1} + \varepsilon_{Sn} k_{ISCn} + \varepsilon k_{bS1} + \varepsilon_{Sn} k_{bSn}\right) P_S + \left(\varepsilon_{T1} \tau_T^{-1} + \varepsilon_{Tn} k'_{ISCn}\right) P_T \\ \dfrac{dP_T}{dt} = \left(\varepsilon k_{ISC1} + \varepsilon_{Sn} k_{ISCn}\right) P_S - \left(\varepsilon_{T1} \tau_T^{-1} + \varepsilon_{Tn} k'_{ISCn} + \varepsilon_{T1} k_{bT1} + \varepsilon_{Tn} k_{bTn}\right) P_T \end{cases}$$

where the parameters (rate constants and relative populations) are given in Fig. 2. To simplify the equation, we let

$$\begin{aligned} k_{ISC} &= k_{ISC1} + (\varepsilon_{Sn}/\varepsilon) k_{ISCn} \\ k_{bS} &= k_{bS1} + (\varepsilon_{Sn}/\varepsilon) k_{bSn} \\ k_T &= \varepsilon_{T1} \tau_T^{-1} + \varepsilon_{Tn} k'_{ISCn} \\ k_{bT} &= \varepsilon_{T1} k_{bT1} + \varepsilon_{Tn} k_{bTn} \end{aligned}$$

where the relative populations are

$$\begin{aligned} \varepsilon &= \frac{k_a k_{Sn1}}{k_a k_{Sn1} + k_a k_{S1n} + \tau_S^{-1} k_{Sn1}} \\ \varepsilon_{Sn} &= \frac{k_a k_{S1n}}{k_a k_{Sn1} + k_a k_{S1n} + \tau_S^{-1} k_{Sn1}} \\ \varepsilon_{T1} &= \frac{k_{Tn1}}{k_{Tn1} + k_{T1n}} \\ \varepsilon_{Tn} &= 1 - \varepsilon_{T1} = \frac{k_{T1n}}{k_{Tn1} + k_{T1n}} \end{aligned}$$

We than have

$$\begin{cases} \dfrac{dP_S}{dt} = -\varepsilon (k_{ISC} + k_{bS}) P_S + k_T P_T \\ \dfrac{dP_T}{dt} = \varepsilon k_{ISC} P_S - (k_T + k_{bT}) P_T \end{cases}$$

When photobleaching is slow, we have the first order approximation of the eigenvalues

$$\begin{cases} -\lambda_1 = -(\varepsilon k_{ISC} + k_T) - \dfrac{\varepsilon^2 k_{bS} k_{ISC} + k_{bT} k_T}{\varepsilon k_{ISC} + k_T} \\ -\lambda_2 = -\dfrac{\varepsilon k_{bS} k_T + \varepsilon k_{bT} k_{ISC}}{\varepsilon k_{ISC} + k_T} \end{cases}$$

and the solutions at the initial conditions $P_S(0) = P_0$ and $P_T(0) = 0$ are

$$\begin{cases} P_S(t) = P_0 \dfrac{1}{\lambda_1 - \lambda_2}\left[(k_T + k_{bT} - \lambda_2)e^{-\lambda_2 t} - (k_T + k_{bT} - \lambda_1)e^{-\lambda_1 t}\right] \\ P_T(t) = P_0 \dfrac{\varepsilon k_{ISC}}{\lambda_1 - \lambda_2}\left(e^{-\lambda_2 t} - e^{-\lambda_1 t}\right) \end{cases}$$

Let

$$\alpha = \lambda_1 - (\varepsilon k_{ISC} + k_T) = \dfrac{\varepsilon^2 k_{bS} k_{ISC} + k_{bT} k_T}{\varepsilon k_{ISC} + k_T}$$

$$\beta = \lambda_2 = \dfrac{\varepsilon k_{bS} k_T + \varepsilon k_{bT} k_{ISC}}{\varepsilon k_{ISC} + k_T}$$

$$k = \lambda_1 - \lambda_2 = (\varepsilon k_{ISC} + k_T) + (\alpha - \beta)$$

and the survival probability of fluorophores under quasi-continuous illumination is

$$R_{cont}(t) = \left(1 + \dfrac{k_{bT} - \alpha}{k}\right)e^{-\beta t} - \left(\dfrac{k_{bT} - \alpha}{k}\right)e^{-\alpha t} e^{-(\varepsilon k_{ISC} + k_T)t}$$

Let

$$\delta = \dfrac{k_{bT} - \alpha}{k} = \dfrac{\varepsilon k_{ISC}(k_{bT} - \varepsilon k_{bS})}{(\varepsilon k_{ISC} + k_T)^2}$$

We then have

$$R_{cont}(t) = e^{-\beta t}\left[(1 + \delta) - \delta e^{-kt}\right]$$

With resonant scanning, Let a single exposure time-span be $t_E$ and a dark time-span be $t_D$. If, 1) $t_D$ is much longer than the triplet-state lifetime; and 2) Photobleaching from $T_n$ dominates over that from $T_1$ (therefore, photobleaching is negligible during $t_D$), then the fluorophore survival probability after $m$ scans with imaging time $t = m(t_E + t_D)$ is

$$R_{scan}(t) = \left\{e^{-\beta t_E}\left[(1 + \delta) - \delta e^{-kt_E}\right]\right\}^m \approx \exp\left(-\eta\left[\beta - \delta(1 - e^{-kt_E})/t_E\right]\cdot t\right)$$

where $\eta = t_E/(t_E + t_D)$ is the scanning duty cycle.

Fluorescence gain of splitting one $t_E$ into $t_E/n$ is

$$\rho(n; \delta) = \dfrac{\left[R_{cont}(t_E/n)\right]^n}{R_{cont}(t_E)} = \dfrac{\left[1 + \delta - \delta\exp(-kt_E/n)\right]^n}{1 + \delta - \delta\exp(-kt_E)}$$

*Photobleaching from the first triplet state ($T_1$) is not significant*

If photobleaching from the first triplet-state ($T_1$) dominated, it would primarily occur during $t_D$. The triplet-state accumulation in $t_E$ would be

$$P_T(t_E) = P_0 \dfrac{\varepsilon k_{ISC}}{k} e^{-\beta t_E}\left(1 - e^{-kt_E}\right)$$

With long enough $t_D$, the photobleached population from $T_1$ in each scan would be

$$B_{T1}(t_E) = \frac{k_{bT1}}{k_{bT1} + \tau_T^{-1}} P_T(t_E)$$

Therefore, the ratio of the photobleached population between with $n$ divided exposure $t_E/n$ and one single continuous $t_E$ is

$$r(n) \approx \frac{n \cdot B_{T1}(t_E/n)}{B_{T1}(t_E)} \approx \frac{n(1-e^{-kt_E/n})}{1-e^{-kt_E}}$$

Note that $r(n)$ is always greater than one. Therefore, if photobleaching from $T_1$ dominated, faster scanning speed would cause more severe photobleaching.

*Sample preparation and labeling*

We used rat and mouse heart samples fixed in 10% formalin and embedded in paraffin, and HeLa cells fixed with 4% paraformaldehyde (cultures at ~50% confluence). Heart thin sections of ~4 µm were embedded in paraffin following standard procedures. Sections were then mounted on slides and heated in an oven at 58°C for 2 hours. After heating, samples were deparaffinized with 100% xylene for 3 × 5 minutes with agitation, placed in decreasing concentrations of ethanol (100%, 95%, 85% and 70%) for 5 minutes two times each, and finally equilibrated in PBS. To achieve antigen unmasking, samples were immersed in a citrate based antigen unmasking solution (Vector Labs, catalog #H-3300; 3.75 ml in 400 ml distilled $H_2O$), microwaved using 7 cycles (2 minutes each) of high heat, and cooled for 1 minute between the heating cycles. After antigen retrieval (heart sections) or fixation (HeLa cells), samples were washed in PBS three times 5 minutes each, equilibrated in 5% NGS-1% BSA-PBS for 30 minutes at room temperature (RT) to block non-specific labeling, incubated with primary antibodies at appropriate dilutions in 0.5% NGS-0.1% BSA-PBS overnight at 4 °C in a humidity chamber, rinsed with PBS three times for 5 minutes, equilibrated to block again for 1 hour at RT (only for heart sample), incubated with the secondary antibody in 0.5% NGS-0.1% BSA-PBS for 60 minutes at RT, washed s three times for 5 minutes in PBS, and finally mounted using ProLong® Gold (Life Technologies, USA) for imaging. Blocking and antibody dilution buffers for HeLa cells immunolabeling contained 0.5% Triton X-100 for permeabilization.

Primary and secondary antibody final concentrations:

<u>HeLa cells:</u>

Sample #1: Anti-ANT pAb 5 µg/ml (Sigma Aldrich, USA, Cat#SAB2105530) and Atto647N goat anti-rabbit IgG 6.7µg/ml (Active Motif Cat#15048).

<u>Heart sections:</u>

Sample #2: Anti-Cytochrome c mAb 5 µg/ml (Abcam, Cat#ab110325) and Atto647 goat-anti-mouse 1 µg/ml (Sigma Aldrich, Cat#50185)

Sample #3. Anti-VDAC mAb 5 µg/ml (Abcam Cat#ab14734) and Oregon Green 488 goat anti-mouse IgG 6.7µg/ml (Molecular Probes® Cat#O-6380).

*Supplementary figures*

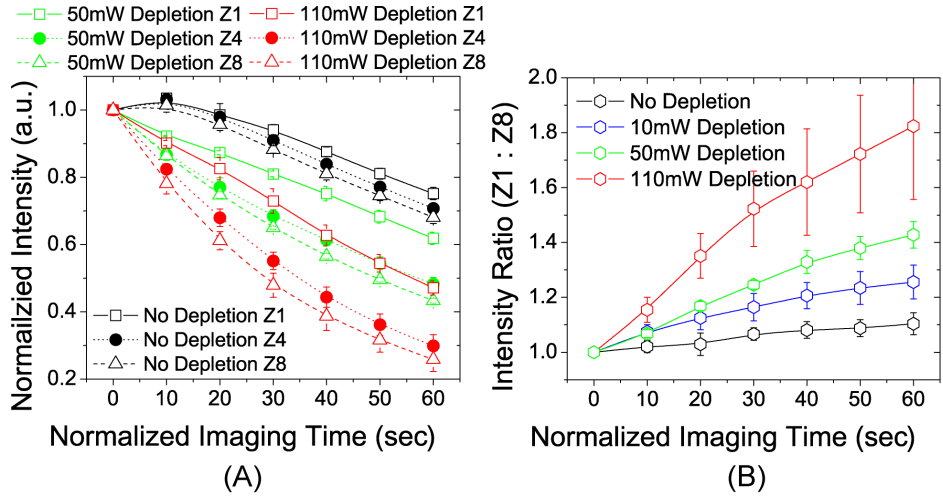

Supplementary Fig S1. Photobleaching rates in Sample #2 with varying depletion laser power. (**A**) Normalized Image intensity decay due to photobleaching as a function of illumination (measured by normalized imaging time). Higher scanning speed (lower zoom) results in slower decay. (**B**) Intensity ratio between Zoom 1 and Zoom 8 is greater than one, and increases with illumination does and with growing depletion laser power. All lines are a guide for the eye.

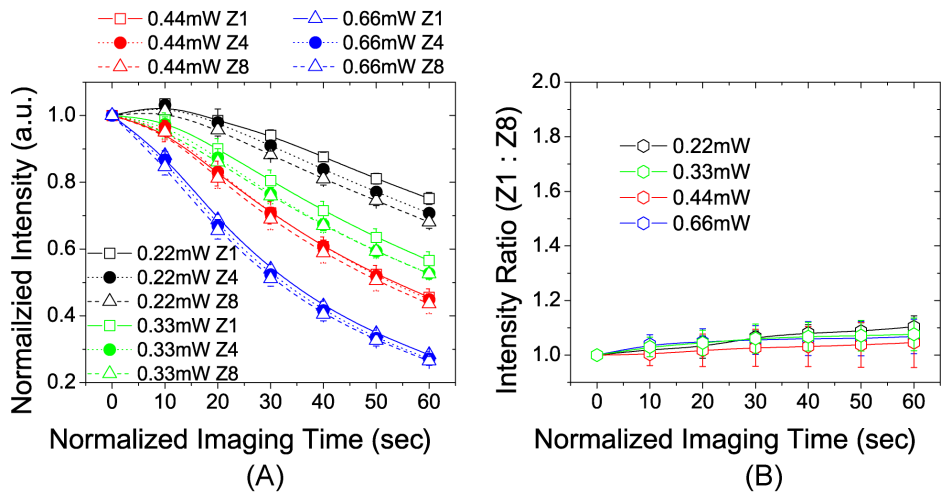

Supplementary Fig. S2. Photobleaching in Sample #2 without depletion (i.e., in regular confocal condition). (A) Image intensity decay curves. (B) Intensity ratio between Zoom 1 and Zoom 8 as function of illumination dose. At high excitation power, photobleaching rate is comparable to that in STED microscopy. Normalized image intensity is similar at all 3 zooms. Difference is at most ~15%. All lines are a guide for the eye.

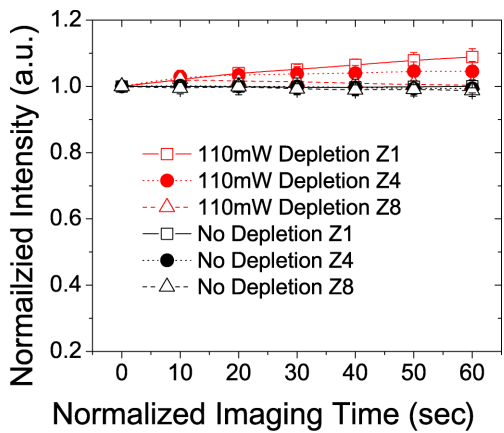

Supplementary Fig S3. Photobleaching caused by depletion laser only (red curves) in Sample #2. Image intensity slightly increased with time, meaning that the depletion laser alone do not cause photobleaching. Black curves represent background photobleaching caused by the low power (16 µW) excitation laser, which is almost zero. See main text for details. Lines are a guide for the eye.

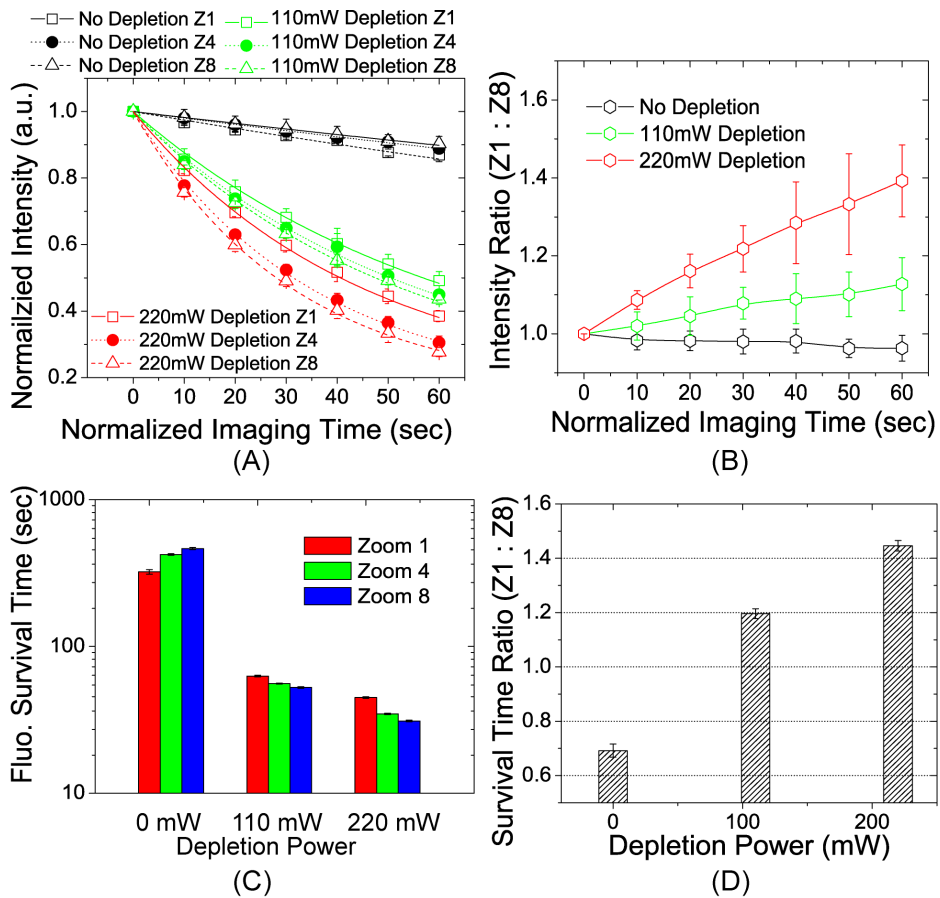

Supplementary Fig.S4. Photobleaching rates in Sample #3 with varying depletion laser power. (A) Image intensity decay data points were fitted to Eq. (3). (B) Intensity ratio between Zoom 1 and Zoom 8 increases with higher illumination dose and with growing depletion laser power. Lines are a guide for the eye. (C) Fluorophore survival time with different depletion laser power at different zooms. Lower zooms (faster scanning speed) have longer survival time when depletion is not zero. (D) Ratio of survival time between Zoom 1 and Zoom 8 is greater than one and increases with growing depletion power.

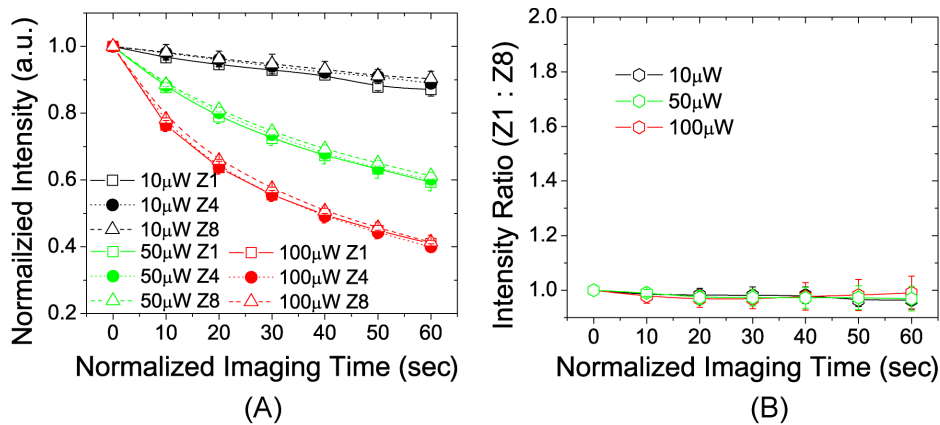

Supplementary Fig.S5. Photobleaching in Sample #3 under regular confocal condition. (A) Image intensity decay curves. (B) Intensity ratio between Zoom 1 and Zoom 8 as function of illumination dose. At high excitation power photobleaching rate is comparable to that in STED microscopy, but photobleaching rates at different zooms are almost the same. All lines are a guide for the eye.

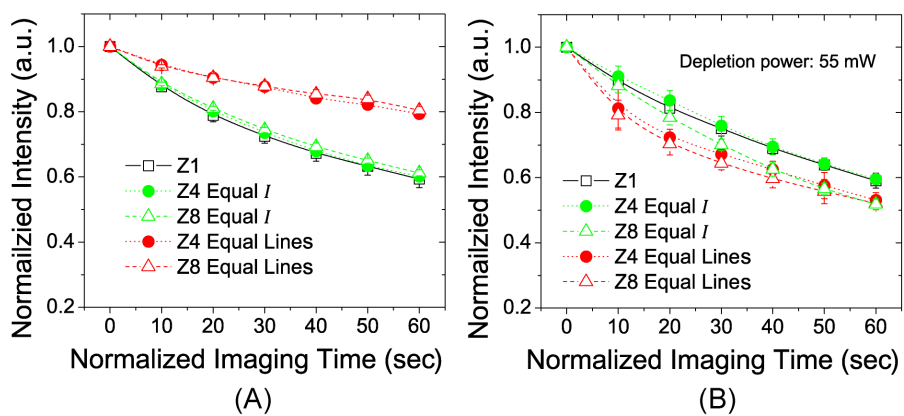

Supplementary Fig S6. Comparison of photobleaching under equal-lines and equal-irradiance (equal I) conditions in Sample #3. Zoom 1 data (black) is used as a reference. Green data points were recorded under equal-irradiance condition (i.e., using the same excitation irradiance as Zoom 1). Red data points were under equal-lines condition, using lower excitation irradiance. (A) In regular confocal microscopy, equal-lines condition has slower photobleaching. (B) In STED microscopy, the depletion power was kept same as in reference to maintain resolution, and equal-lines condition has faster photobleaching. All lines are a guide for the eye.

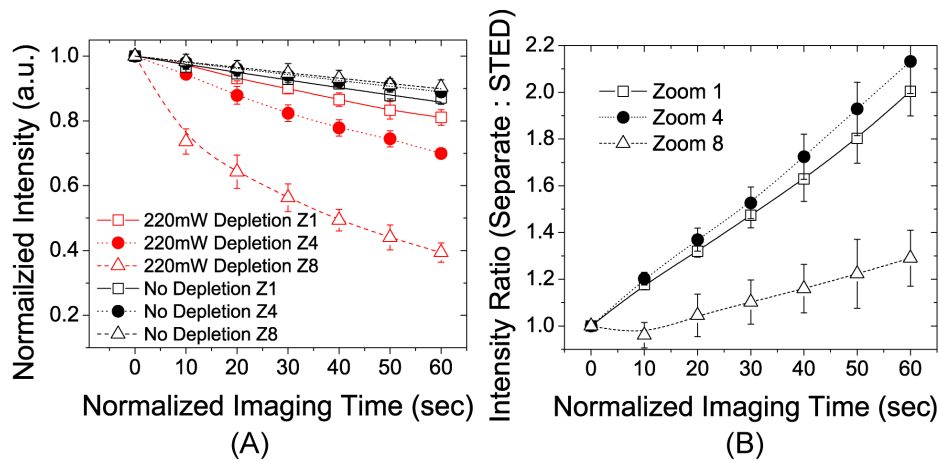

Supplementary Fig S7. (A) Image intensity decay with separated excitation and depletion illumination in Sample #3. The red data points resulted from using both excitation laser beam (485 nm, 50 µW) and depletion laser beam (592 nm, 220 mW), but they were on separately (i.e., never switched on at same time). Black curves are with excitation laser on only. Significant photobleaching was caused by depletion laser alone, because red curves are much lower than black ones. At higher zoom (slower scanning speed), depletion-only photobleaching is more severe. (B) Photobleaching caused by separated excitation and depletion illumination is slower than STED photobleaching. The image intensity ratio between separate illumination and STED (50 µW excitation and 220 mW depletion being on together) is almost always greater than one, and keeps increasing with time. All lines are a guide for the eye.

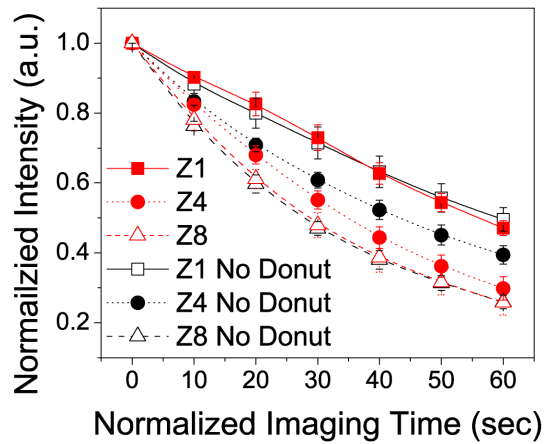

Supplementary Fig S8. Image intensity decay with 0.22 mW excitation power and 110 mW depletion laser power in Sample #2. The red curves show results with doughnut-shaped depletion laser beam, whereas the black curves are results with Gaussian depletion beam. The difference is small, especially at Zoom 1 and Zoom 8.